\documentclass[aps,prd,twocolumn,floatfix,nofootinbib,superscriptaddress,showkeys]{revtex4-2}

\usepackage{amsmath,amsfonts,extarrows,amssymb,mathrsfs,times,bm}
\usepackage{extarrows}
\usepackage{graphicx}
\usepackage{float}
\usepackage{graphicx,epstopdf}
\usepackage{dcolumn}
\usepackage{exscale}
\usepackage{relsize}
\usepackage{mathtools}
\usepackage{mhchem}
\usepackage[usenames,dvipsnames]{pstricks}
\usepackage{subfigure}
\usepackage{epsfig}
\usepackage{pst-grad} 

\usepackage[colorlinks=true,breaklinks=true,linkcolor=blue,citecolor=blue,urlcolor=blue]{hyperref}
\newcommand{\nn}{\nonumber}

\usepackage{ulem}
\usepackage{color}

\newcommand{\delred}[1]{{\color{red}{\ifmmode\text{\sout{\ensuremath{#1}}}\else\sout{#1}\fi}}}

\begin{document}

\title{Deflection of charged signals in a dipole magnetic field in Kerr background}

\author{Zonghai Li}
\affiliation{School of Physics and Technology, Wuhan University, Wuhan, 430072, China}

\author{Junji Jia}
\email[Corresponding author:~]{junjijia@whu.edu.cn}
\address{Department of Astronomy \& MOE Key Laboratory of Artificial Micro- and Nano-structures, School of Physics and Technology, Wuhan University, Wuhan, 430072, China}

\date{\today}

\begin{abstract}
This paper investigates charged particle deflection in a Kerr spacetime background with a dipole magnetic field, focusing on the equatorial plane and employing the weak field approximation. We employ the Jacobi-Randers metric to unify the treatment of the gravitational and electromagnetic effects on charged particles. Furthermore, we utilize the Gauss-Bonnet theorem to calculate the deflection angle through curvature integrals. The difference between the prograde and retrograde deflection angles is linked to the non-reversibility of metrics and geodesics in Finsler geometry, revealing that this difference can be considered a Finslerian effect. We analyze the impact of both gravitomagnetic field and dipole magnetic field on particle motion and deflection using the Jacobi-Randers magnetic field. The model considered in this paper exhibits interesting features in the second-order approximation of ($M/b$). When $q\mu=2MaE$, the Jacobi-Randers metric possesses reversible geodesics, leading to equal prograde and retrograde deflection angles. In this case, the gravitomagnetic field and dipole magnetic field cancel each other out, distinguishing it from scenarios involving only the gravitomagnetic field or the dipole magnetic field. We also explore the magnetic field's impact on gravitational lensing of charged particles.
\end{abstract}

\maketitle

\section{Introduction}
In usual electrodynamics in flat spacetime, the dipole magnetic field under spherical coordinates $(r, \theta, \phi)$ is given by~\cite{refC56.Jackson}:
\begin{align}
	\mathbf{B}=\frac{\mu}{r^{3}}\left(2\cos \theta\frac{\partial}{\partial r}+ \frac{1}{r} \sin \theta\frac{\partial}{\partial \theta}\right),
	\label{c6.eq:flatmag}
\end{align}
where $\mu$ is the magnetic dipole moment. In 1930, St\"{o}rmer systematically investigated the trajectories of charged particles under the influence of this magnetic field~\cite{refc6.Stormer1930}. Subsequently, the study of the trajectory of charged particles in the dipole magnetic field has been known as the St\"{o}rmer's problem, which is of great significance for understanding the Earth's magnetic field and related phenomena (such as auroras)~\cite{refc6.Stormer1955}. In recent years, interest in this problem has mainly been in the field of nonlinear analysis~\cite{refc6.Dragt-dipole,refc6.Markellos-dipole,WL-dipole,LL-dipole}. In curved spacetimes, the form of the dipole magnetic field will have to be modified due to the spacetime curvature.

The investigation of electromagnetic fields in curved spacetime can be carried out using a conventional approximation method. This method treats the electromagnetic field as a perturbation that is influenced by spacetime, but whose effect on spacetime is ignored. The key aspects of this approach include the modification of Maxwell's equations to account for the influence of curved spacetime, and the exclusion of the electromagnetic field's energy-momentum tensor from gravitational field equations due to the negligible impact of the former on spacetime. With this approximation, the electromagnetic field in curved spacetime has been widely studied~\cite{Ginzburg,Anderson-Cohen,refc6.Cohen-Wald,refc6.Wald-uniform,refc6.Prasanna-EM1,Petterson-SD,Petterson-KD,refc6.Kerr-dipole}. Particularly interesting is the case of a dipole magnetic field, as the magnetic field of dense stars such as neutron stars is usually approximated by a dipole, and, for black holes, their dipole magnetic field can be generated by current loop on the equatorial plane. The model of a dipole magnetic field in curved (for example, Schwarzschild~\cite{Petterson-SD} and Kerr~\cite{Petterson-KD}) spacetimes can be used to study related issues\cite{Preti2004,Bakala2010,refc6.Shapiro-NS,refc6.Sengupta-NS,refc6.Rezzolla-NS, refc6.Potekhin-NS,refc6.Beskin-NS,KD-chaotic}.

Analyzing the behavior of charged particles under the dual influence of electromagnetic and gravitational fields holds paramount astronomical significance. One notable example is its pivotal role in comprehending the origins, propagation, and distinctive features of cosmic rays, which serve as essential messengers in multi-messenger astronomy. In consideration of the significant application of gravitational lensing in astronomy, this paper embarks on a theoretical exploration of the gravitational deflection effect in charged particles within a dipole magnetic field against the background of spacetime around a rotating black hole. This investigation serves as a continuation of the same theme in the Kerr-Newman spacetime~\cite{LJ-KN2021} and Schwarzschild spacetime with an existing dipole magnetic field~\cite{LWJ-SD2022}. In this series of studies, we employ the Jacobi-Randers metric and Gauss-Bonnet theorem. The Jacobi-Randers metric allows for a uniform treatment of the influences of electromagnetic and gravitational fields on charged particles. Utilizing the Gauss-Bonnet theorem, we calculate the deflection angle based on curvatures, a methodology initially introduced by Gibbons and Werner~\cite{GW2008} and subsequently adopted and expanded upon by various authors, such as those of Refs.~\cite{Werner2012,OIA2017,GGK-High,LHZ2020,LZA2020,LJepjc,HuangSC2023}. In the example in this article, the motion of the charged particles is affected by two effects related to the prograde or retrograde motion directions. The first is the gravitomagnetic effect inherent in rotating space-time, and the second is the external dipole magnetic field. The effect of these two parts makes the model have a rich and interesting gravitational lensing.

This paper is structured as follows. In Section \ref{Preliminaries}, we outline the fundamental prerequisites for the upcoming sections. This includes detailing the methodology for computing the deflection angle using the Jacobi-Randers metric and Gauss-Bonnet theorem, elucidating the magnetic field definition based on the Jacobi-Randers metric, and exploring the Finslerian effect on the deflection angle. In Section \ref{Kerr-dipole}, we introduce the Kerr-dipole background, derive the orbit equations, and determine the deflection of charged particles within the framework of the weak-field approximation. In Section \ref{sec:dis}, we delve into the Finslerian effect, analyze the impact of magnetic fields on deflection, and explore relevant aspects of gravitational lensing. Finally, we conclude the paper in Section \ref{conclusion}. Throughout this paper, we adopt units such that $G = c = 1$, and the spacetime signature is $(-,+,+,+)$.

\section{Jacobi-Randers metric in curved spacetime}\label{Preliminaries}

\subsection{Jacobi Metric for a charged particle}

In this paper, we consider the motion and deflection of charged particles in a 4-dimensional spacetime with an electromagnetic field $(M,g_{\mu\nu},A_\mu)$. The line element with the coordinate system $(t,x^i)(i=1,2,3)$ can be written as,
\begin{align}
d s^2=g_{t t}(x) d t^2+2 g_{t i}(x) d t d x^i+g_{i j}(x) d x^i d x^j.
\end{align}
In addition, we also assume that the electromagnetic gauge field $A_{\mu}$ is independent of time. 

Charged particles in $(M,g_{\mu\nu},A_\mu)$ are subject to both gravitational and electromagnetic fields, and typically, their motion does not follow geodesics. Due to the conservation of particle energy, their trajectories can be described as geodesics of the $3D$ Jacobi metric, following Maupertuis' principle of least action. In other words, we can use the Jacobi metric to unify the dual effects of the gravitational and electromagnetic fields. The Jacobi metric for a particle 
 (mass $m$, energy $E$, and charge $q$) moving in $(M,g_{\mu\nu},A_\mu)$ is given by~\cite{Chanda2019b}
\begin{align}
\label{fensleranders}
F(x,dx)=d\rho=\sqrt{\alpha_{ij}dx^idx^j}+\beta_idx^i,
\end{align}
with
\begin{subequations}\label{sub-fensleranders}
\begin{align}
\label{fensleranders1}
&\alpha_{ij}=\frac{\left(E+qA_t\right)^2+m^2{g}_{tt}}{-{g}_{tt}}\left({g}_{ij}-\frac{{g}_{ti}{g}_{tj}}{{g}_{tt}}\right),\\
\label{fensleranders2}
&\beta_i=qA_i-\left(E+qA_t\right)\frac{g_{ti}}{g_{tt}}.
\end{align}
\end{subequations}
Here, $F$ is a Finsler metric of Randers type, where $\alpha_{ij}$ is a Riemannian metric and $\beta_i$ is a one-form, satisfying positivity and convexity
\begin{align}
\label{FCpc}
|\beta|=\sqrt{\alpha^{i j} \beta_{i} \beta_{j}}<1.
\end{align}

For easier reference in the following sections, we designate the Riemannian part $\alpha_{ij}$ in the (Jacobi-)Randers metric \eqref{fensleranders} as the (Jacobi-)Randers-Riemann metric, characterized by the line element
\begin{align}
dl^2=\alpha_{ij}dx^idx^j,
\end{align}
with the unit velocity vector 
\begin{align}
e^i=\frac{dx^i}{dl}.
\end{align}

\subsection{Equations of Motion, Randers Magnetic Field, and Gauge Invariance}

In this subsection, based on the work of Gibbons et al. \cite{Gibbons-Inv}, we discuss the general Randers metric, and these considerations naturally apply to the Jacobi-Randers metric given by Eqs. \eqref{fensleranders}- \eqref{sub-fensleranders}.
The Lagrangian describing the motion of a free particle in the Randers metric space is given by
\begin{align}
\label{Lagrangian-RF}
\mathcal{L}=F(x,\dot{x})=\sqrt{\alpha_{ij}\dot{x}^i\dot{x}^j}+\beta_i\dot{x}^i,
\end{align}
where a dot represents derivatives with respect to any parameter. Now we choose $\dot{x}^i=dx^i/dl=e^i$,  
then the equation of motion can be obtained by Euler-Lagrange equations, as follows~\cite{Perlick-Inv,OIA2017}
 \begin{align}
 \label{motion-equation}
\frac{de^i}{d l}+\Gamma_{j k}^i e^je^k=\Tilde{F}^i_k e^k.
\end{align} 
In the above, $\Gamma_{j k}^i$ denotes the Christoffel symbols associated with the Randers-Riemann metric $\alpha_{ij}$, and
\begin{align}
\label{emf}
\Tilde{F}_{ij}=\beta_{j;i}-\beta_{i;j}=\beta_{j,i}-\beta_{i,j},\quad \Tilde{F}_k^i=\alpha^{ij}\Tilde{F}_{jk},
\end{align}
where, the semicolon ``$;$'' signifies the covariant derivative with respect to the metric $\alpha_{ij}$, while the comma ``$,$'' denotes partial derivatives.

Using the same parametrization, the motion equation of a charged particle with mass $\mathcal{M}$ and charge $\mathcal{Q}$ in a magnetic field characterized by the magnetic potential $A_i$ can be expressed as follows~\cite{Gibbons-Inv}
\begin{align}
\label{motion-equation-mag}
\frac{de^i}{d l}+\Gamma_{j k}^i e^je^k=\frac{\mathcal{Q}}{\sqrt{2\mathcal{M}\mathcal{E}}}F^i_k e^k,
\end{align} 
where the energy $\mathcal{E}$ is
\begin{align}
\mathcal{E}=\frac{m}{2}\alpha_{ij}\frac{dx^i}{d\tau}\frac{dx^j}{d\tau},
\end{align} 
with $\tau$ representing proper time, and
\begin{align}
F_{ij}=A_{j;i}-A_{i;j}=A_{j,i}-A_{i,j},\quad F_k^i=\alpha^{ij}F_{jk}.
\end{align}
 
Comparing Eq. \eqref{motion-equation} and Eq. \eqref{motion-equation-mag}, we can establish the following analogy
\begin{align}
\label{analogous}
\quad \beta_i= \frac{\mathcal{Q}}{\sqrt{2\mathcal{M}\mathcal{E}}}A_i.
\end{align}
Based on the analogous relation \eqref{analogous}, one can employ the Randers data $(\alpha_{ij},\beta_i)$ to establish the following magnetic field (referred to as the Randers magnetic field)~\cite{Gibbons-Inv}
\begin{align}
	\label{c3.reducemag}
	\Tilde{B}^i=\alpha^{ij}(^*d\beta)_j=\frac{1}{2}\epsilon^{ijk}\Tilde{F}_{jk}.
\end{align}
In the above, the symbol ``$*$'' represents the Hodge star operator, and $\epsilon^{ijk}=\frac{1}{\sqrt{\alpha}}\varepsilon^{ijk}$ is the Levi-Civita tensor, where $\alpha\equiv \det (\alpha_{ij})$ represents the determinant of $\alpha_{ij}$, and $\varepsilon^{ijk}$ denotes the Levi-Civita symbol.

Regarding the Jacobi-Randers metric presented in Eqs. \eqref{fensleranders}- \eqref{sub-fensleranders}, a natural consideration pertains to its gauge invariance under the electromagnetic gauge transformation,
\begin{align}
\label{gau-trans}
A_{\mu}\rightarrow A_{\mu}'=A_{\mu}+\partial_{\mu}\Psi,
\end{align}
where $\Psi$ represents any scalar field. The answer is negative. However, this does not alter the kinematics since the motion equation \eqref{motion-equation} satisfies electromagnetic gauge invariance. More generally, Eq. \eqref{motion-equation} remains invariant under the following transformation~\cite{Gibbons-Inv,Perlick-Inv}
\begin{align}
\label{GII}
\beta_i\to \beta'_i=\beta_i+\partial_i\Psi.
\end{align}

\subsection{Equation of trajectory}

In Boyer-Lindquist coordinates $(t, r, \theta, \phi)$, the metric for a stationary and axisymmetric spacetime can be expressed as follows
\begin{align}
\label{SAS-le}
d s^2=&g_{t t}(r,\theta) d t^2+2 g_{t \phi}(r,\theta) d t d \phi+g_{rr}(r,\theta) dr^2\nn\\
&+g_{\theta\theta}(r,\theta) d\theta^2+g_{\phi\phi}(r,\theta) d\phi^2.
\end{align}
Due to the time translation and axial symmetry, for any equatorial motion ($\theta=\pi/2,x=(r,\phi)$) there exist two motion constants, energy $E$ and angular momentum $L$. They are expressed in terms of the asymptotic velocity $v$ as follows:
\begin{align}
\label{conservedel}
E=\frac{m}{\sqrt{1-v^2}},\quad L=\frac{mb}{\sqrt{1-v^2}}=Eb,
\end{align}
where $b$ is the impact parameter.

The orbit equation for particles in the equatorial plane can be directly derived from the Jacobi-Randers metric $(\alpha_{ij},\beta_i)$. Introducing the inverse radial coordinate as $u = 1/r$, the orbit equation is given as follows~\cite{LWJ-SD2022}
\begin{align}
\label{S2.orbit-eq}
\left(\frac{du}{d\phi}\right)^2=u^4\frac{\alpha_{\phi\phi}\left[\alpha_{\phi\phi}-\left(L-\beta_\phi\right)^2\right]}{\alpha_{rr}\left(L-\beta_\phi\right)^2}.
\end{align}

In general, solving the above equations for a given spacetime is challenging. However, in the weak-field regime of interest in this paper, it can be solved perturbatively. The commonly used boundary conditions are
\begin{align}
(1)~u(0)=0,\quad \text{or}\quad (2)~u'(\frac{\pi}{2})\equiv\frac{du}{d\phi}(\frac{\pi}{2})=0.
\end{align}

\subsection{Deflection angle using the Gauss-Bonnet theorem}
The Gauss-Bonnet theorem is a fundamental theorem in differential geometry, which establishes a connection between curvature and the topological properties of surfaces. The deflection angle characterizes the degree to which the trajectory of a particle is bent while propagating in a gravitational field and serves as a fundamental quantity in gravitational lensing phenomena. By applying the Gauss-Bonnet theorem to a two-dimensional Riemannian space, where the trajectory of the particle corresponds to spatial curves within that space, an exact expression for the deflection angle in terms of spatial curvature can be obtained. This method was introduced by Gibbons and Werner~\cite{GW2008} and has since become a valuable tool for studying gravitational lensing.

When dealing with the particle's trajectory in a Randers space, a more technical treatment is required to apply the Gauss-Bonnet theorem. Two popular approaches have been widely adopted to transform the problem into a Riemannian space. One approach involves using Osculating Riemannian metric, introduced by Werner~\cite{Werner2012}, and the other employs Randers-Riemann metric $\alpha_{ij}$, introduced by Ono, Ishihara, and Asada ~\cite{OIA2017}. This paper adopts the method in that work, which is based on the motion equation Eq. \eqref{motion-equation}. According to Eq. \eqref{motion-equation}, the motion of a free particle in $(M^3,F=\sqrt{\alpha_{ij}dx^idx^j}+\beta_idx^i)$ can be equivalent to the motion of a charged particle in $(M^3,\alpha_{ij},\Tilde{B}^i)$, where the Randers magnetic field $\Tilde{B}^i$ is defined by Eq. \eqref{c3.reducemag}. 

\begin{figure}[htp!]
\centering
\includegraphics[width=8.0cm]{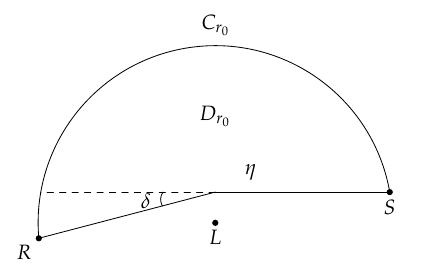}
\caption{A non-singular region $D_{r_0}\subset(M^2,\alpha_{ij})$, bounded by particle ray $\eta$ and a circular curve $C_{r_0}$ (defined by $r=r_0$). The particle travels from the source point $S$ to the receiver point $R$ and undergoes deflection due to the gravitational lens $L$. The deflection angle is denoted as $\delta$.\label{Figure}}
\end{figure}

Below, we will utilize the method in Ref. \cite{OIA2017} to examine particle deflection in the equatorial plane. We assume that the two-dimensional background space $(M^2, \alpha_{ij})$ is asymptotic Euclidean and is described by coordinates $(r, \phi)$. The particle travels from the source point $S$ to the receiver point $R$ and undergoes deflection due to the lens $L$. We consider a non-singular region $D_{r_0}\subset(M^2,\alpha_{ij})$, which is enclosed by the particle trajectory $\eta$ and an auxiliary curve $C_{r_0}$ (defined by $r=r_0$), as illustrated in Fig. \ref{Figure}. 

By applying the Gauss-Bonnet theorem to the region $D_{r_0}$, with the assumption that $S$ and $R$ are located in the asymptotic region, the deflection angle $\delta$ can be expressed in terms of the Gaussian curvature $K$ of $\alpha_{ij}$ and the geodesic curvature $k_g$ of the particle's trajectory. This expression is given as follows (for details see Ref.~\cite{LWJ-SD2022})
\begin{align}
\label{DEF-Gg}
\delta=\underbrace{-\iint_{D_{r_0}(r_0\to \infty)} K d S}_{\delta_{G}}+\underbrace{\int_{S}^{R} k_g dl}_{\delta_{g}},
\end{align}
Notably, the contribution of the geodesic curvature term, denoted as $\delta_{g}$, arises from the typically non-zero Jacobi-Randers magnetic field $\Tilde{B}^i$.

Gaussian curvature $K$ of $\alpha_{ij}$ can be calculated by~\cite{Werner2012}
\begin{align}
\label{G-cur}
K=\frac{1}{\sqrt{\alpha}}\bigg[\frac{\partial}{\partial \phi}\left(\frac{\sqrt{\alpha}}{\alpha_{r r}} \Gamma_{r r}^{\phi}\right)-\frac{\partial}{\partial r}\left(\frac{\sqrt{\alpha}}{\alpha_{r r}} \Gamma_{r \phi}^{\phi}\right)\bigg].
\end{align}
To compute the geodesic curvature $k_g$, one can refer to OIA~\cite{OIA2017} and utilize the following formula
\begin{align}
\label{geod-mag01}
k_g=&\left[-\frac{\beta_{\phi,r}^{3D}}{\sqrt{\alpha^{3D} \alpha_{3D}^{\theta\theta}}}\right]_{\theta=\pi/2},
\end{align}
where the scripts $3D$ denotes 3-dimensional Jacobi-Randers data $(\alpha_{ij},\beta_i)$ with coordinates $(r,\theta,\phi)$.
With the spacetime line element~\eqref{SAS-le}, the Jacobi-Randers-Riemann metric $\alpha_{ij}$ given by Eq. \eqref{fensleranders1} is diagonal, thus the geodesic curvature simplifies to
\begin{align}
\label{g-cur}
k_g=-\frac{1}{\sqrt{\alpha}}\beta_{\phi,r}
\end{align}

When using the Gauss-Bonnet theorem to calculate higher-order deflection angles through Eq. \eqref{DEF-Gg}, there are certain iteration rules to be followed. In the coordinates $(r,\phi)$, to calculate the $n$-th order deflection angle, information from the $(n-1)$-th order particle trajectory and the $(n-2)$-th order deflection angle is essential~\cite{GGK-High}. For the orbit satisfying $u(0)=0$, the deflection angle \eqref{DEF-Gg} can be refined as 
\begin{subequations}\label{GB-DEf1}
\begin{align}
&\delta_{G}=-\int_{0}^{\pi+\delta^{[n-2]}} \int_{0}^{u^{[n-1]}} K\sqrt{\alpha} du d\phi,\label{GB-DEf1a}\\
&\delta_{g}=\int_{0}^{\pi+\delta^{[n-2]}} \left(k_g \frac{ dl}{d\phi}\right)d\phi.\label{GB-DEf1b}
\end{align}
\end{subequations}
Here, and in the subsequent expressions, the superscript $[n]$ indicates that a quantity is accurate to the $n$-th order, while the superscript $(n)$ indicates a quantity at the $n$-th order. For instance, the second-order deflection angle can be denoted as $\delta^{[2]}=\delta^{(1)}+\delta^{(2)}$. In Sec. \ref{GBDA}, we will utilize Eq. \eqref{GB-DEf1} to compute the deflection angle of charged particles in the Kerr-dipole background. Given the intricate interplay of gravitational, gravitomagnetic, and magnetic lensing, it is imperative to investigate higher-order deflections. Therefore, we will compute deflection angles up to the third order.

Although the second boundary condition $u'(\frac{\pi}{2})=0$ will not be used in this paper, its corresponding deflection angle formula is still worth mentioning, as follows,
\begin{subequations}
\label{GB-DEf2}
  \begin{align}
&\delta_{G}=-\int_{-\frac{1}{2}\delta^{[n-2]}}^{\pi+\frac{1}{2}\delta^{[n-2]}} \int_{0}^{u^{[n-1]}} K\sqrt{\alpha} du d\phi,\\
&\delta_{g}=\int_{-\frac{1}{2}\delta^{[n-2]}}^{\pi+\frac{1}{2}\delta^{[n-2]}}\left(k_g \frac{ dl}{d\phi}\right)d\phi.
\end{align}
\end{subequations}
It is easy to see that when the order of calculation is less than or equal to two, the expressions corresponding to the two boundary conditions are the same.

\subsection{Divergence in Deflection Angles Between Prograde and Retrograde Particles: A Finsler Geometry Perspective}\label{FinsEffe}
A Finsler metric $F(x,y)(x\in M,y\in T_xM)$ is said to be reversible or symmetric if it equals its retrograde Finsler metric $\bar{F}(x,y)=F(x,-y)$~\cite{Ohta}. For example, the Riemannian metric is reversible,
\begin{align}
\bar{F}(x,y)&=\sqrt{\alpha_{ij}(-y^i)(-y^j)}\nn\\
&=\sqrt{\alpha_{ij}y^iy^j}=F(x,y).
\end{align}
For the Randers metric, we have  
\begin{align}\label{eq:rmfinal}
	\bar{F}(x,y)=&\sqrt{\alpha_{ij}(-y^i)(-y^j)}+\beta_i(-y^i)\nn\\
=&\sqrt{\alpha_{ij}y^iy^j}-\beta_iy^i\nn\\
=&\sqrt{\bar{\alpha}_{ij}y^iy^j}+\bar{\beta}_iy^i,
\end{align}
where
\begin{align}
\label{dubDEF}
&\bar{\alpha}_{ij}=\alpha_{ij},\quad \bar{\beta}_i=-\beta_i.
\end{align}
Therefore, a Randers metric is reversible if and only if $\beta_i=0$, meaning it is a Riemannian metric. The non-Riemannian Randers metric is non-reversible. In fact, Randers introduced the Randers metric as a generalization of the Riemannian metric by considering asymmetry (non-reversibility)~\cite{Randers}.

The non-reversibility of Finsler metric leads to the non-reversibility of geodesics. For the Finsler metric $F(x,y)$, the inverse curve of its geodesic (i.e., orbit tracing backward along the original curve) is not necessarily its geodesic. However, we have the fact that the curve $\eta$ is the geodesic of $F(x,y)$ if and only if its inverse curve $\bar{\eta}$ is the geodesic of $\bar F(x,y)$~\cite{Ohta}. If a Finsler metric $F(x,y)$ possesses the property that the inverse curves of all its geodesics are also its geodesics, we refer to $F(x,y)$ as having reversible geodesics~\cite{Crampin}. Clearly, a reversible Finsler metric must have reversible geodesics. However, it is also possible for a non-reversible Finsler metric to have reversible geodesics. Crampin~\cite{Crampin} has shown that, for the Randers metric, it has reversible geodesics if and only if $\beta_i$ is closed (equivalently, $\Tilde{F}_{ij}=0$). From Eq. \eqref{motion-equation}, it is evident that the geodesics of $F$ in this case coincide with the geodesics of $\alpha_{ij}$. 

In the context of this paper, we employ the Finsler-Randers metric $F(x,y)$ to study the deflection angles of geodesic trajectories of particles from a source $S$ to a receiver $R$. The deflection angles of its inverse curves are investigated using the inverse metric $\bar F(x,y)$. 
In conclusion, the difference in deflection angles between prograde and retrograde particles is a consequence of the non-reversibility of the Finsler metric, or more precisely, the non-reversibility of geodesics. Thus, we can regard the difference in deflection of particles in the prograde and retrograde directions as a Finslerian effect. It is a common phenomenon found in various scenarios, such as the deflection of particles in rotating spacetime~\cite{LJepjc} and the deflection of charged particles in a dipole magnetic field~\cite{LWJ-SD2022}. 

It's worth noting that in situations where the metric is non-reversible but the geodesics are reversible, the deflection angles for prograde and retrograde trajectories remain the same. This is unlike the Sagnac effect, which can also be considered a Finslerian effect~\cite{Masiello01,Masiello02}. The Sagnac effect is solely determined by the non-reversibility of the optical-Randers metric. In other words, analogous to the Aharonov-Bohm effect, the Sagnac effect is determined by the non-zero potentials $\beta_i$.

Finally, 
based on Eq.~\eqref{dubDEF}, the retrograde deflection angle can be obtained from the prograde deflection angle by adding a negative sign to the appropriate parameter (such that Eq.~\eqref{dubDEF} holds).

\section{Deflection angle of charged particles in Kerr spacetime with a dipole magnetic field}\label{Kerr-dipole}
\subsection{Kerr spacetime with a dipole magnetic field}

The Kerr metric~\cite{Kerr.bh} describes the spacetime geometry surrounding a rotating black hole with mass $M$ and angular momentum per unit mass $a$. In the Boyer-Lindquist coordinates $(t,r,\theta,\phi)$, its line element is
\begin{eqnarray}
\label{spacetime.geometry}
ds^{2}&=&-\left(1-\frac{2 M r}{ \Sigma}\right) d t^{2}-\frac{4 M r a\sin ^{2} \theta}{ \Sigma}d t d \phi+\frac{\Sigma }{\Delta} d r^{2} \nn\\
&&+\Sigma d \theta^{2}+\frac{\left(r^{2}+a^{2}\right)^{2}-\Delta a^{2}\sin^2\theta}{\Sigma} \sin ^{2} \theta d \phi^{2},
\end{eqnarray}
where
\begin{eqnarray}
&&\Sigma=r^{2}+a^{2} \cos ^{2} \theta,\quad\Delta=r^{2}-2 M r+a^{2}. \nn
\end{eqnarray}

In Kerr spacetime, the corresponding electromagnetic potential of a dipole magnetic field can be written as~\cite{Petterson-KD,KD-chaotic}
\begin{eqnarray}
\label{ele-mag-gf}
A_{t}&=&-\frac{3 a \mu}{2 \Sigma \zeta^{2}}\bigg\{\left[r(r-M)+\left(a^{2}-M r\right) \cos ^{2} \theta\right]\nn\\
&&\times\frac{1}{2 \zeta} \ln \left(\frac{r-M+\zeta}{r-M-\zeta} \right)-\left(r-M \cos ^{2} \theta\right)\bigg\} ,\nn\\
A_{\phi}&=&-\frac{3 \mu \sin ^{2} \theta}{4 \Sigma \zeta^{2}}\bigg\{(r-M) a^{2} \cos ^{2} \theta+r\left(r^{2}+M r+2 a^{2}\right)\nn\\
&&-\frac{1}{2 \zeta}\left[r\left(r^{3}-2 M a^{2}+a^{2} r\right)+\Delta a^{2} \cos ^{2} \theta\right] \nn\\
&&\times\ln \left(\frac{r-M+\zeta}{r-M-\zeta}\right)\bigg\}
\end{eqnarray}
where $\zeta=\sqrt{M^2-a^2}$, and $\mu$ is the magnetic dipole moment. 

\subsection{Kerr-dipole Jacobi-Randers metric and its magnetic field}

The $3D$ Kerr-dipole Jacobi-Randers metric $(\alpha^{3D}_{ij},\beta^{3D}_i)$ and its reversibility analysis is given by Appendix~\ref{3data}. Here, we focus on the equatorial plane, and the corresponding Jacobi-Randers metric can be obtained from Eqs. \eqref{appRFa}- \eqref{appRFc} by setting $\theta=\pi/2,d\theta=0$, as follows
\begin{subequations}
\label{Randers-KD}
\begin{align}
\label{Randers-KD-a}
&\alpha_{ij}dx^idx^j=\gamma r^2\left(  \frac{dr^2}{\Delta}+\frac{\Delta}{r^2-2Mr}d\phi^2\right),\\
&\beta_\phi= \frac{1}{8 \zeta^3r\left(r-2M \right)}\bigg\{-16MaEr+3 q \mu \times\nn\\
&~~\bigg[4r\zeta^3+2(M-r)\zeta r^2+\Delta r^2\log \left(\frac{r-M+\zeta}{r-M-\zeta}\right)\bigg]\bigg\},
\end{align}
\end{subequations}
where
\begin{align}
\label{Randers-KD-c}
\gamma=&-m^2+\frac{r}{r-2M}\bigg\{E+\frac{3aq\mu}{2\zeta^2r^2}\nn\\
&\bigg[r-\frac{r(r-M)\log \left(\frac{r-M+\zeta}{r-M-\zeta}\right)}{2\zeta}\bigg]\bigg\}^2.
\end{align}

This paper does not aim for exact solutions. For the convenience of presenting approximate orders,
we denote the mass $M$, spacetime spin $a$ and magnetic dipole $\mu$ as $f_1,~f_2$ and $f_3$ respectively with proper power of a scale factor $\varepsilon$, as follows
\begin{align}
M=f_1 \varepsilon,\quad a=f_2\varepsilon,\quad \mu=f_3 \varepsilon^2.
\end{align}

Utilizing Eq.~\eqref{KdJRmag}, we expand the Jacobi-Randers magnetic field on the equatorial plane to third order, yielding 
\begin{subequations}
\label{Randers-MF}
\begin{align}
\Tilde{B}^{r}=&\Tilde{B}^{\phi}=0,\\
\Tilde{B}^{\theta}=&\frac{\varepsilon^2}{(E^2-m^2)^{3/{2}}}\left(\frac{qf_3}{r^4}-\frac{2f_1f_2E}{r^4}\right)\nn\\
&+\bigg[\frac{6m^2f_1f_2E}{r^5}-\frac{1}{r^5}(2m^2+E^2)qf_3\bigg]\nn\\
&\times\frac{f_1\varepsilon^3}{(E^2-m^2)^{5/{2}}}+\mathcal{O}(\varepsilon^4).
\end{align}
\end{subequations}

Initially, we observe that the magnetic field exclusively exhibits a non-zero component in the $\theta$ direction. This implies that the magnetic field is orthogonal to the equatorial plane, allowing particles to remain stable within it. In addition, when considering the second-order magnetic field, intriguing characteristics become apparent. It comprises the gravitomagnetic field $\Tilde{\mathbf{B}}_{GM}$ and the dipole magnetic field $\Tilde{\mathbf{B}}_M$, given by
\begin{align}
&\Tilde{\mathbf{B}}_{GM}=-\frac{f_1f_2E\varepsilon^2}{r^4(E^2-m^2)^{3/{2}}}\frac{\partial}{\partial \theta},\\
&\Tilde{\mathbf{B}}_{M}=\frac{qf_3\varepsilon^2}{r^4(E^2-m^2)^{3/{2}}}\frac{\partial}{\partial \theta}.
\end{align}
Without loss of generality, we assume both $a=f_2\varepsilon$ and $q\mu=qf_3\varepsilon^2$ to be non-negative. Consequently, the gravitomagnetic field is oriented perpendicular to the equatorial plane and directed upwards, reducing the particle's deflection angle. In contrast, the dipole magnetic field is also oriented perpendicular to the equatorial plane but directed downwards, leading to an increase in the particle's deflection angle. The effect of the total magnetic field on the deflection angle depends on whether the dipole magnetic field or the gravitomagnetic field dominates, determining whether it increases or decreases.

In a special case, when the gravitomagnetic field and the dipole magnetic field cancel each other out, i.e., $qf_3=2Ef_1f_2$ (or $q\mu=2EaM$), the total magnetic field makes no contribution to the deflection angle. The situation corresponds to Kerr-dipole Jacobi-Randers metric having reversible geodesics (in the sense of second-order approximation), which can be seen from Eq.~\eqref{rev.effect}. It is worth noting that when only the gravitomagnetic field or only the dipole magnetic field is present, the Jacobi-Randers metric does not possess reversible geodesics, and the magnetic field always contributes to the deflection angle. Examples of this scenario include the motion and deflection of neutral particles in the Kerr field (with only the gravitomagnetic field)~\cite{LJepjc} or charged particles in the Schwarzschild+dipole field (with only the dipole magnetic field)~\cite{LWJ-SD2022}. 

When considering the third-order contributions, the discussion above becomes ineffective. Substituting $qf_3=2f_1f_2E$ into Eq. \eqref{Randers-MF}, it is found that the magnetic field is 
\begin{align}
&\Tilde{\mathbf{B}}=-\frac{2f_1^2f_2E\varepsilon^3}{(E^2-m^2)^{3/2}}\frac{\partial}{\partial \theta}+\mathcal{O}(\varepsilon^4).
\end{align}
Clearly, in this case, the magnetic field reduces the deflection of the particle trajectory. 

Quantitative investigations into the deflection of particle trajectories by magnetic fields necessitate the computation of deflection angle. In the following sections, we shall undertake this task to determine the extent to which magnetic field influences particle trajectories. 

\subsection{The weak-filed trajectory}\label{litaibai}

By substituting the Jacobi-Randers data $(\alpha_{ij},\beta_i)$ from Eq. \eqref{Randers-KD} into Eq. \eqref{S2.orbit-eq}, we obtain the orbit equation for a charged particle in Kerr spacetime with a dipole magnetic field as follows
\begin{align}
\label{KD-orbit}
\left(\frac{du}{d\phi}\right)^2=&\frac{1}{b^2}-u^2+2 u\left(\frac{1}{v^2}-1+b^2 u^2\right)\frac{f_1 \varepsilon}{b^2}\nn\\
&+
\bigg[b f_2^2 u^2\left(3-2 b^2 u^2\right) v\nn\\
&-4 f_1f_2u+\frac{2 q f_3u}{E}\bigg]\frac{\varepsilon^2}{b^3 v}+\mathcal{O}\left(\varepsilon^3\right),
\end{align}
where we have used Eq. \eqref{conservedel}.

Consider the second-order expansion of the trajectory $u = u(\phi)$
\begin{align}
\label{partray3rd}
u^{[2]} = u^{(0)} + u^{(1)} + u^{(2)}.
\end{align}
Substituting this into the orbit equation Eq.~\eqref{KD-orbit} and utilizing the initial condition $u(0) = 0$, we iteratively solve from lower to higher orders. We can determine the second-order solution for the motion as
\begin{subequations}
\label{subpartray3rd}
\begin{align}
u^{(0)}=&\frac{\sin \phi}{b},\\
u^{(1)}=&\frac{(1-\cos\phi)\left(1-v^{2} \cos\phi\right)}{b^2 v^{2}}f_1\varepsilon,\\
u^{(2)}=&\left(\frac{q}{E}f_3-2 f_1 f_2\right) \frac{\varepsilon^2(1-\cos \phi)}{b^3 v}+\frac{ \sin^3\phi}{2 b^3}f_2^2\varepsilon^2\nn\\
&-\frac{f_1^2\varepsilon^2}{8 b^3 v^{2}} \bigg\{6\left(4+v^{2}\right) \phi-16\left(1+v^{2}\right) \sin\phi\nn\\
&+\left[-8+7 v^{2}+3 v^{2} \cos(2\phi)\right] \tan\phi\bigg\}\cos\phi.
\end{align}
\end{subequations}

\subsection{Higher order deflection angle}\label{GBDA}

From Eqs. \eqref{Randers-KD-a} and \eqref{Randers-KD-c}, it is easy to see that the Jacobi-Randers-Riemann metric $\alpha_{ij}$ is asymptotically Euclidean,
\begin{align}
dl^2=\alpha_{rr}dr^2+\alpha_{\phi\phi}d\phi^2\to E^2v^2(dr^2+r^2d\phi^2).
\end{align}
Therefore, the deflection angle can be calculated using Eq. \eqref{DEF-Gg}, specifically Eq. \eqref{GB-DEf1}, as the particle trajectories given by Eqs. \eqref{partray3rd}- \eqref{subpartray3rd} satisfy the condition $u(0)=0$. The purpose here would be to calculate the deflection angle to the third order, Eq. \eqref{GB-DEf1} can be rewritten as 
\begin{align}
\label{eq:d4totalsum}
\delta^{[3]}=\delta_{G}^{[3]}+\delta_{g}^{[3]},
\end{align}
where
\begin{subequations}\label{def-ang-KD}
\begin{align}
\label{Gdef-ang-KD}
&\delta_{G}^{[3]}=-\int_{0}^{\pi+\delta^{[1]}} \int_{0}^{u^{[2]}} K\sqrt{\alpha} du d\phi,\\
\label{gdef-ang-KD}
&\delta_{g}^{[3]}=\int_{0}^{\pi+\delta^{[1]}} \left(k_g \frac{ dl}{d\phi}\right)d\phi.
\end{align}
\end{subequations}. 

Among them, the second-order particle trajectory $u^{[2]}$ is determined by Eq. \eqref{partray3rd} with Eq. \eqref{subpartray3rd}. To compute the third-order deflection angle, we also need to know the Gaussian curvature $K$ of $\alpha_{ij}$, the geodesic curvature $k_g$ of the particle ray $\eta$, and the first-order deflection angle $\delta^{[1]}$.

Substituting the Jacobi-Randers-Riemann metric $\alpha_{ij}$ from Eq. \eqref{Randers-KD} into Eq. \eqref{G-cur}, we can calculate the Gaussian curvature as follows
\begin{align}
\label{KD-G-cur}
-K\sqrt{\alpha}=&\left(1+\frac{1}{v^2}\right) f_1 \varepsilon+\left(1-\frac{4}{v^4}+\frac{6}{v^2}\right) f_1^2 u \varepsilon^2\nn\\
&+3\bigg[\left(\frac{1}{2}+\frac{4}{v^6}-\frac{10}{v^4}+\frac{15}{2 v^2}\right) f_1^3-\frac{3 f_2 f_3 q}{2 Ev^2}\nn\\
&+\left(1+\frac{1}{v^2}\right) f_1 f_2^2\bigg] u^2 \varepsilon^3+\mathcal{O}\left(\varepsilon^4\right).
\end{align}

Comparing Eqs. \eqref{geod-mag01} and \eqref{KdJRmag}, we observe the relationship between the magnetic field and the geodesic curvature
\begin{align}
k_g=\left[\sqrt{\alpha^{3D}_{\theta\theta}}\Tilde{B}^{\theta}_{3D}\right]_{\theta=\pi/2}=\Tilde{B}^{\theta}\left[\sqrt{\alpha^{3D}_{\theta\theta}}\right]_{\theta=\pi/2}.
\end{align}
Thus, by utilizing Eqs. \eqref{Randers-MF} and \eqref{appRFa}, we can determine the geodesic curvature. Alternatively, we can directly substitute the values of $(\alpha_{ij},\beta_i)$ provided by Eq. \eqref{Randers-KD} into Eq. \eqref{g-cur} to obtain it. The resulting expression for the geodesic curvature is found to be
\begin{align}
\label{KD-g-cur}
 k_g \left(\frac{dl}{d\phi}\right)&=\left(\frac{qf_3}{E}-2 f_1 f_2\right)\frac{\varepsilon^2 \sin\phi}{b^2 v}\nn\\
 &+\bigg[2\left(2+\frac{1}{v^2}\right) \frac{f_1 f_3 q}{E}-\left(3+\frac{1}{v^2}\right) 4 f_1^2 f_2\nn\\
 &-2 f_1\left(4 f_1 f_2-\frac{f_3 q}{E}\right) \cos\phi\bigg]\sin^2 (\frac{\phi}{2}) \frac{\varepsilon^3}{b^3 v} \nn\\
&+\mathcal{O}\left(\varepsilon^4\right),
\end{align}
where we have employed the particle's orbit given by Eq.~\eqref{partray3rd} with Eq.~\eqref{subpartray3rd}.

From Eqs. \eqref{KD-G-cur} and \eqref{KD-g-cur}, it is evident that the first-order deflection angle is solely influenced by the Gaussian curvature component, and there is no contribution from both the gravitomagnetic field and dipole magnetic field. By using Eq. \eqref{KD-G-cur} in conjunction with the zero-order particle trajectory, which is given by $u^{[0]}=\sin\phi/b$, this can be readily obtained. Interestingly, this scenario is entirely identical to the Schwarzschild spacetime case with no electromagnetic field, yielding the following result
\begin{align}
\label{eq:d2total}
\delta^{[1]}=\int_{0}^{\pi} \int_{0}^{u^{[0]}} \left(1+\frac{1}{v^2}\right) f_1 \varepsilon du d\phi=\frac{4f_1\varepsilon}{b}.
\end{align}

Incorporating the third-order Gaussian curvature \eqref{KD-G-cur}, the second-order particle trajectory \eqref{partray3rd}-\eqref{subpartray3rd}, and the first-order deflection angle \eqref{eq:d2total} into Eq. \eqref{Gdef-ang-KD} and performing the necessary integration, the third-order deflection angle contributed by the Gaussian curvature can be expressed as follows
\begin{align}
\label{eq:G4res}
\delta_{G}^{[3]}=2f_1&\left(1+\frac{1}{v^2}\right)\frac{\varepsilon}{b}+\frac{3 \pi f_1^2}{4 }\left(1+\frac{4}{v^2}\right)\frac{\varepsilon^2}{b^2}\nn\\
&+\bigg[\frac{2f_1^3}{3}\left(5+\frac{45}{v^2}+\frac{15}{v^4}-\frac{1}{v^6}\right)\nn\\
&+2f_1\left(f_2^2-\frac{\pi f_1f_2}{v}+\frac{\pi f_3 }{2v}\frac{q}{E}\right)\nn\\
&\times\left(1+\frac{1}{v^2}\right)-\frac{2 f_2 f_3 }{ v^2}\frac{q}{E}\bigg]\frac{\varepsilon^3}{b^3}.
\end{align}
The deflection angle determined by the geodesic curvature can be obtained by substituting Eqs. \eqref{KD-g-cur} and \eqref{eq:d2total} into Eq. \eqref{gdef-ang-KD}. The result up to the third order is
\begin{align}
\label{eq:g4res}
\delta_g^{[3]}=&2\left(\frac{f_3 q}{ E }-2 f_1 f_2\right)\frac{\varepsilon^2}{b^2v}+\bigg[\frac{ f_1 f_3 q}{ E}\left(\frac{3}{2}+\frac{1}{v^2}\right)\nn\\
&-{2 f_1^2 f_2}\left(2+\frac{1}{v^2}\right)\bigg]\frac{\pi\varepsilon^3}{b^3v}.
\end{align}

Finally, utilizing Eqs.~\eqref{eq:G4res} and \eqref{eq:g4res}, we obtain the total third-order deflection angle $\delta^{[3]}=\delta_{G}^{[3]}+\delta_{g}^{[3]}$, as presented in Eqs. \eqref{deflection-angle}-\eqref{sub-def-ang} below, which includes the result for retrograde particle rays.

\section{Discussion of results\label{sec:dis}}
\subsection{Finslerian effect}
For the particle ray reaching the detector from the opposite direction, the deflection angle is calculated by $F$'s inverse metric $\bar{F}$, according to Sec. \ref{FinsEffe}. Observing the Randers data in Eq. \eqref{Randers-KD}, we notice that the transformation from $(\alpha_{ij},\beta_i)$ to $(\bar{\alpha}_{ij},\bar{\beta}_i)$ is given by $a \rightarrow -a,q\mu \rightarrow -q\mu$. Consequently, the deflection angle for the retrograde trajectory can be obtained by applying the same transformation to the prograde deflection angle. By expressing the results for both directions in a unified form, we can write the deflection angle that applies to both trajectory rotations as
\begin{align}
\label{deflection-angle}
\delta^{[3]}=\delta^{(1)}+\delta^{(2)}+\delta^{(3)},
\end{align}
where
\begin{subequations}
\label{sub-def-ang}
\begin{align}
\delta^{(1)}=& 2\left(1+\frac{1}{v^2}\right)\frac{M}{b}, \\
\label{sub-def-ang2}
\delta^{(2)}=& \frac{3 \pi}{4}\left(1+\frac{4}{v^2}\right) \frac{M^2}{b^2}+\frac{2s}{b^2v}\left(\frac{ q \mu}{E }-2Ma\right),\\
\label{sub-def-ang3}
\delta^{(3)}=& \frac{2}{3}\left(5+\frac{45}{v^2}+\frac{15}{v^4}-\frac{1}{v^6}\right) \frac{M^3}{b^3}\nn\\
&+\frac{s\pi}{b^3v}\left[\frac{1}{2}\left(5+\frac{4}{v^2}\right) \frac{ q \mu M}{E }-2\left(3+\frac{2}{v^2}\right)a M^2\right]\nn\\
&+2 \left(1+\frac{1}{v^2}\right)\frac{a^2 M}{b^3}-\frac{2 a q \mu}{E b^3v^2}.
\end{align}
\end{subequations}
Among them, $s=+1$ and $s=-1$ represent the deflection angles of the prograde and retrograde particle rays, respectively.

\subsection{Effect of magnetic field}

When $q\mu\to 0$, the result given in Eqs. \eqref{deflection-angle}-\eqref{sub-def-ang} becomes the deflection angle of massive neutral particles in Kerr spacetime~\cite{Jia-Kerr2020}. When $a\to 0$, it reduces to the deflection angle of charged particles in Schwarzschild spacetime with a dipole magnetic field~\cite{LWJ-SD2022}. 

In the current case, we see from Eq. \eqref{sub-def-ang2} that the dipole magnetic field starts to appear from second order in the form of $\sim sq\mu/(Evb^2)$, which can be traced back to Eq. \eqref{eq:g4res} in the geodesic curvature contribution. 
The sign of this term depends on the sign of the charge $q$ and the direction of the magnetic dipole $\mu$. For a prograde (or retrograde) trajectory, the above term implies that when the charge is positive $(q>0)$ and the magnetic field in the equatorial plane is downward $(\mu>0)$, it contributes positively (or negatively) to the deflection angle so that the trajectory is bent more (or less) to the lens. 

In Ref. \cite{LWJ-SD2022} it was argued that the effect of the magnetic field at the second order is quantitatively comparable to that of the spacetime spin under replacement $q\mu\longleftrightarrow -2EM a$. From \eqref{sub-def-ang2} we see that this is exactly the case here. However, from Eq. \eqref{sub-def-ang3} this equivalence is seen broken immediately from the third order, which shows the fundamental difference between the Lorentz force and the gravitational interaction. If the parameters are such that $q\mu =2EM a$, then we see that the second-order influence from the dipole magnetic field and the gravitomagnetic field cancel each other. This cancellation can be traced back to Eq. \eqref{Randers-MF}, where the Jacobi-Randers magnetic field becomes zero at the second order when this relation is set. A similar situation has been observed in the Kerr-Newman spacetime~\cite{LJ-KN2021}. A more dramatic case would be to choose  
\begin{align}
    \frac{q\mu}{E}=-\frac{3\pi sv}{8}\left( 1+\frac{4}{v^2}\right) M^2+2Ma
\end{align} 
such that the entire second-order deflection is canceled out. Note that for black hole spacetime, since $|a|\leq M$, cancellation of the entire 
second order term was not possible if there was only the spacetime spin. This becomes possible only when the magnetic effect is taken into account, because all parameters $q,~\mu$ and $E$ can have wide ranges of variation.

From Eq. \eqref{sub-def-ang3}, we observe that the surface (Gaussian) curvature contribution of the magnetic field to the deflection manifests from one order higher than the Lorentz force term, that is, from the third order in the form of $qa\mu/(Ev^2b^3)$. This term also couples the magnetic field with the spacetime spin, and is not trajectory direction dependent. Note however this is not the only third-order contribution from the magnetic field.

\subsection{Gravitational lensing}

With the deflection angles known, we can establish the lensing equation for signals originating from the source at radius $r_s$ and solve for the apparent angles of these signals, in the eyes of an observer located at radius $r_d$. 
Previously, using the deflection up to the second order, the apparent angles have also been solved to this order for the Kerr spacetime without the magnetic field \cite{ZhangFanJia}. 

Now because of the extra term added to the spacetime spin term in the second order deflection angle, i.e. $-2Ma\to -2Ma+q\mu/E$, the apparent angle to the second order in the current case, denoted as $\theta_{Km}$ can be directly obtained by substituting $a\to a-q\mu/(2EM)$ into the apparent angle, Eq. (5.9) of Ref. \cite{ZhangFanJia}. The result is
\begin{align}
\label{eq:thetakerr}
\theta_{Km}=\frac{b_{0s}}{r_d}+\frac{b_{1s}}{r_d}
+\mathcal{O}\left( \varepsilon^3\right),
\end{align}
where
\begin{subequations}
\label{eq:kbkdefs}
\begin{align}
b_{0s}=&\frac{\varphi_0 r_d r_s}{2( r_d+r_s)}( \sqrt{ 1+\eta } -s ),\\
b_{1s}=&\frac{\eta [ 8s(\frac{q\mu}{EM}-2 a)v+3M \pi ( 4+v^2) ]}{32( 1+v^2)\sqrt{
1+ \eta }(\sqrt{
1+ \eta }-s)} ,\label{eq:kb1}\\
\eta =&\frac{8M( r_d +r_s )}{\varphi^2_0 r_d r_s } ( 1+ \frac{1}{v^2} ). \label{eq:kbkdefs3}
\end{align}
\end{subequations}

We then observe that both the spacetime spin and the magnetic dipole will not affect the image's apparent angles at the leading order. They appear simultaneously in the second order. From Eq. \eqref{eq:kb1} we see that they appear in the same combination as in the second order of the deflection angle, i.e., Eq. 
\eqref{sub-def-ang3}. This also suggests that the effects of the spin and dipole magnetic field on the image's apparent angles can cancel (or enhance) each other in this order. Moreover, the result \eqref{eq:thetakerr} as a perturbative result, also requires the smallness of $b_{1s}$ compared to $b_{0s}$. This in turn implies that when $q\mu/(EM)\gg b\gg 2a$, this apparent angle formula will break down. This case indeed corresponds to the situation that the Lorentz force is repulsive from the center and much stronger than gravity, and the deflection angle \eqref{sub-def-ang} becomes invalid. 

\section{Conclusion}\label{conclusion}

In this paper, we have studied the deflection of charged particles in a rotating spacetime with a dipole magnetic field. Our approach involves the use of the Jacobi-Randers metric to provide a unified treatment of the gravitational and electromagnetic influences on these particles. The dipole magnetic field and the gravitomagnetic field are merged into the Jacobi-Randers magnetic field. The difference between prograde and retrograde deflection angles is linked to the non-reversibility of metrics and geodesics in Finsler geometry, revealing this difference as a Finslerian effect. In the model considered in this paper, under the second-order approximation, the Jacobi-Randers metric exhibits reversible geodesics, resulting in identical prograde and retrograde deflection angles, which differs from cases where only gravitomagnetic field or dipole magnetic fields are present.

The deflection angle is computed to the third order of $M/b$. It is observed that the dipole magnetic field's influence on deflection is at least of second order, consistent with the order at which spacetime spin (or the gravitomagnetic field) emerges. Interestingly, at this second order, when the parameters are adjusted to satisfy $q\mu\sim 2EaM$, the dipole field will balance the gravitomagnetic field. Consequently, at this order, both the deflection and the apparent angles in gravitational lensing will remain unaffected by either of these fields. As mentioned earlier, at this point, they are direction-independent. This balance, however, breaks down at the third order, and in general, achieving full balance is impossible.

The apparent angles of the gravitational lensing images are also affected by the magnetic interaction from the second order. The cancellation of the magnetic and spacetime spin effect also happens at the same value of the parameters $(q,\,\mu,\,E,\,M,\,a)$, as that in the deflection angle. 

\acknowledgements

This work is supported by grants from NNSF and MST China. 

\appendix
\begin{widetext}
\section{Three-dimensional Kerr-dipole Jacobi-Randers metric and an analysis of its reversibility}\label{3data}
The $3D$ Randers metric $(\alpha^{3D}_{ij},\beta^{3D}_i)$ can be obtained by substituting the spacetime metric \eqref{spacetime.geometry} and electromagnetic potential \eqref{ele-mag-gf} into Eq.\eqref{sub-fensleranders}, as follows
\begin{align}
\label{appRFa}
&\alpha^{3D}_{ij}dx^idx^j=\gamma\Sigma\left(  \frac{dr^2}{\Delta}+d\theta^2+\frac{\Delta\sin^2\theta}{\Delta-a^2\sin^2\theta}d\phi^2\right),\\
\label{appRFb}
&\beta^{3D}_\phi= \frac{\sin^2\theta}{8 \zeta^3\left(\Delta-a^2 \sin^2\theta\right)}\bigg\{-16MaEr+3 q \mu \bigg[4r\zeta^3+2(M-r)\zeta\Sigma+\Delta\Sigma\log \left(\frac{r-M+\zeta}{r-M-\zeta}\right)\bigg]\bigg\},
\end{align}
where
\begin{align}
\label{appRFc}
\gamma=-m^2+\frac{\Sigma}{\Sigma-2M r}\bigg\{E+\frac{3aq\mu}{2\zeta^2\Sigma}\bigg[r-M\cos^2\theta-\frac{(\Delta-(a^2-Mr)\sin^2\theta)\log \left(\frac{r-M+\zeta}{r-M-\zeta}\right)}{2\zeta}\bigg]\bigg\}^2.
\end{align}
From Eq.\eqref{appRFb}, we find that for the
axis direction $(\theta=0,\pi)$, $\beta^{3D}_i=0$, that is the Kerr-dipole Jacobi-Randers metric is reversible. 
Substituting Eq. \eqref{appRFb} into Eq. \eqref{emf}, we have
\begin{align}
\Tilde{F}^{3D}_{r\phi}=& \frac{\sin^2\theta}{\left(\Delta-a^{2} \sin^2\theta\right)^2} \bigg\{2 a M E\left(r^2-a^2 \cos^2\theta\right) +\frac{q }{8 \zeta^3}\bigg[12(r-M) \zeta \mu\left(r\Sigma-M\Sigma-2 \zeta^2 r\right)\nn\\
&+6 \mu \log \left(\frac{r-M+\zeta}{r-M-\zeta}\right)\left(r \Delta^2-a^2(r \Delta+(r-M) \Sigma) \sin^2\theta\right)\nn \\
& -6 \zeta \mu\left(\Delta-a^{\wedge} 2 \sin^2\theta\right)\left(3 a^2-2\left(M^2+M r-2 r^2\right)+a^2 \cos2\theta]\right)\bigg]\bigg\} ,\\
\Tilde{F}^{3D}_{\theta\phi}=& \frac{\sin2\theta}{\left(\Delta-a^2 \sin^2\theta\right)^2} \bigg\{\frac{3 q \mu }{8 \zeta^3} \bigg[\Delta \log \left(\frac{r-M+\zeta}{r-M-\zeta}\right)\left(\Sigma^2-2 M r\left(r^2+a^2 \cos2\theta\right)\right)\nn\\
&+2 \zeta\left(M\Sigma^2-r\Sigma^2+2 M r^2\left(\Sigma-2 M^2\right)-2 a^2 r\left(\Delta+M(r-2 M) \sin^2\theta\right)\right)\bigg]-2 a M E r \Delta \bigg\}.
\end{align}
For the
axis direction $(\theta=0,\pi)$, $\Tilde{F}_{r\phi}=\Tilde{F}_{\theta\phi}=0$, that is the Kerr-dipole Jacobi-Randers metric has reversible geodesics. Additionally, for the equatorial plane ($\theta=\pi/2$), only $\Tilde{F}_{r\phi}$ non-zero, if we set $\Tilde{F}_{r\phi}=0$ , we obtain
\begin{align}
\label{rev.eq}
\frac{q}{E}=\frac{(8 a M  \zeta^3)/(3\mu)}{2\left(\zeta^2+M^2-3 M r+r^2\right) \zeta+\left[a^2 M-r(r-2 M)^2\right] \log \left(\frac{r-M+\zeta}{r-M-\zeta}\right)}.
\end{align}
For particles moving on a geodesic circular orbit $(r=r_0=\text{constant})$, if they satisfy the above equation, then this circular orbit is reversible. If we only consider the lowest-order effects and expand Eq. \eqref{rev.eq}, we obtain

\begin{align}
\label{rev.effect}
\frac{q}{E}=\frac{2Ma}{\mu}+\mathcal{O}(\frac{1}{r}).
\end{align} 

Subsequently, using Eq. \eqref{c3.reducemag}, we can calculate the Kerr-dipole Jacobi-Randers magnetic field, which leads to the following outcome
\begin{align}
\label{KdJRmag}
\Tilde{\mathbf{B}}_{3D}=\frac{1}{\sqrt{\alpha^{3D}}}\left(\Tilde{F}^{3D}_{\theta\phi}\frac{\partial}{\partial r}-\Tilde{F}^{3D}_{r\phi}\frac{\partial}{\partial \theta}\right),
\end{align}
where 
\begin{align}
\alpha^{3D}=\frac{(\gamma \Sigma)^3)sin^2\theta}{\Delta-a^2\sin^2\theta}.
\end{align}
Along the axis direction $(\theta=0,\pi)$, the magnetic field has only the radial component, so the particles are not affected by the Lorentz force. On the equatorial plane $(\theta=\pi/2)$ the magnetic field has components only in the $\theta$ direction. Consequently, the Lorentz force due to $\Tilde{\mathbf{B}}$ acting on the particle is confined to the equatorial plane, ensuring that the particle remains on this plane. 
\end{widetext}

\end{document}